\documentclass [12pt]
{article}
\usepackage{amsmath}
\usepackage{amsfonts}
\usepackage{feynmf}
\usepackage{feyn}
\newcommand{\bean}{\begin{eqnarray*}}
\newcommand{\eean}{\end{eqnarray*}}
\newcommand{\ed}{\end{document}}

\newcommand{\be}{\begin{equation}}
\newcommand{\ee}{\end{equation}}
\newcommand{\barr}{\begin{array}}
\newcommand{\earr}{\end{array}}
\newcommand{\bea}{\begin{eqnarray}}
\newcommand{\eea}{\end{eqnarray}}

\begin{document}
\title{Solving
field equations in spinor electrodynamics.}
\author{A.V.Bratchikov
 \\ Kuban
State Technological University,\\ 2 Moskovskaya Street, Krasnodar,
  350072, Russia
\\
}
 \date{} \maketitle
\begin{abstract}
Solutions of classical and quantum equations of motion in spinor electrodynamics are constructed within the context of  perturbation theory. The solutions possess a graphical representation in terms of diagrams.
\end{abstract}



\section  {Introduction}
Quantum  spinor electrodynamics describes photons interacting with electrons and positrons. Scattering amplitudes of these interactions can be computed using pertubation theory. Terms of the corresponding series are represented by the Feynman diagrams.

The aim of this paper is to demonstrate a similar 
method for solving  
equations of motion in spinor electrodynamics. 

Let $x=(x^0,x^1,x^2,x^3)
$ be space-time coordinates,$\eta^{\mu\nu}=diag (1,-1,-1,-1)$ the metric tensor, and $\Box=\partial^\mu \partial _\mu.$ 
The basic equations of spinor electrodynamics are defined by \cite {AB},
\begin {gather} 
\Box A_\mu+
j_\mu
=0,\notag \\     \label{b}
  i\bar \psi\gamma ^\mu {\overset {\leftarrow} \partial}_ \mu 
 +e \bar\psi\gamma ^\mu A_\mu + m \bar \psi =0,\\
i\gamma ^\mu \partial_ \mu \psi  
-e \gamma ^\mu A_\mu\psi - m \psi=0. \notag
\end {gather}
Here 
$A_{\mu}(x)$ is the four-vector potential of the electromagnetic field,\, $\psi(x)$ is the Dirac field with charge $e$ and mass $m, \bar \psi(x) 
$ is the Dirac adjoint, $\gamma ^\mu$ are the gamma matrices, $\gamma ^\mu \gamma ^\nu+ \gamma ^\nu \gamma ^\mu =2\eta^{\mu\nu},$ and  
\bean
j_\mu=\frac {e} 2 [\bar \psi, \gamma_\mu \psi]
.
\eean
The fields $ A_\mu,\psi, \bar \psi$ are operator-valued functions in quantum theory and elements of the Berezin algebra in  classical one \cite {GT}.
The classical equations are compatible with the Lorenz gauge condition
\bean \label{D}
\partial_\mu A^\mu=0.
\eean
In what follows $ A_\mu,\psi$ and  $\bar \psi$ can be treated as elements of an arbitrary algebra over ${\mathbb C}.$

In the present paper we find explicit expressions for the electromagnetic and Dirac fields in terms of the corresponding free fields. These expressions are given by power series in charge $e.$ Each term of the series is represented by a sum of diagrams. We ignore the divergences of quantum electrodynamics.
The form of our solution
in quantum case is different from that of ref. \cite {S}.

The paper is organized as follows.  In the next section we review properties of the system of equations (\ref{b}) and rewrite it in a convenient form. 
In Sec.3 we construct a solution for a wide class of nonlinear equations with quadratic nonlinearity and introduce a graphical representation. In Sec.4 we describe the solution of equation (\ref{b}).

\section {Equations of motion in integral form
}

Equations (\ref {b}) can be transformed to their integral forms, see \cite {CG}. Suppose  $A_{0\mu },{\psi}_0$ and $\bar { \psi}_0$ are general solutions of the free equations
\begin {gather*}\label{U}
\Box A_{0 \mu }=0, 
\quad 
i \bar \psi_0\gamma ^\mu {\overset {\leftarrow} \partial}_ \mu+ m \bar \psi_0=0,\quad 
i\gamma ^\mu \partial_ \mu \psi_0-m \psi_0=0.
\end {gather*}
Then equations  (\ref {b}) take the form
\begin {gather} \label{i}
A = A_{0}+ \langle \bar { \psi},  \psi \rangle,\notag \\
\bar { \psi}= \bar { \psi}_0+ \langle   \bar {\psi},A \rangle,\quad
\label{ii} {\psi}= \psi_0+ \langle  A,  \psi \rangle.
\end {gather}
Here $A=A_\mu dx^\mu,$$A_{0}=A_{0\mu }dx^\mu,$ $\langle \bar \psi, \psi \rangle= J_\mu dx^\mu, $
\bean
J_\mu=-\frac 1 {4\pi}\int^t_0\tau d\tau \int_{S} 
j_\mu(t-\tau, x^1+\tau\xi^1, x^2+\tau\xi^2, x^3+\tau\xi^3) d \sigma_\xi, 
\eean 
$\xi^1, \xi^2$ and $\xi^3$ are coordinates on the unit sphere $S$, $\sigma_\xi$ is the area element on $S,$ $t=x^0,$ 
\begin {gather*}
\langle  A,  \psi \rangle  = i e \gamma^0 e^{tK}\int_0^te^{-tK}( \gamma ^\mu 
A_\mu \psi
) dt, \qquad K= \gamma^0 
(\sum_{k=1}^3 \gamma^k \partial_k +im) 
,\\
\langle \bar {\psi},  {A} \rangle= -i e 
 \int_0^t (\Bar {\psi} \gamma ^\mu A_\mu) 
e^{-t
\overset{\leftarrow}{K}}dt e^{t
\overset{\leftarrow}{K}}\gamma^0,\qquad \overset{\leftarrow}{K}= 
(\sum_{k=1}^3 \gamma^k 
\overset{\leftarrow}{\partial}_k +im 
) \gamma^0.
\end{gather*} 

Let 
${\cal A}, \varPsi$ and $\bar \varPsi $ be spaces of 1-forms, the
Dirac spinors and Dirac adjoints, respectively.
Define 
\begin {gather*}
\langle  \psi,  A\rangle=\langle A, \psi \rangle, \quad \langle  \psi, \bar \psi \rangle=
\langle \bar  \psi, \psi  \rangle, \quad\langle  A,\bar \psi\rangle
=\langle \bar \psi,   A \rangle,\\
\langle A, A \rangle =\langle \psi,  \psi \rangle=\langle \bar \psi, \bar \psi \rangle=0.
\end {gather*} 
Then for $\Phi=A+ \psi(x)+ \bar \psi(x)\in {\cal A}\oplus \varPsi\oplus \bar \varPsi$ the system of equations (\ref{ii}) takes the form \bea \label {qu0}
\Phi= \Phi_0+\frac {1} {2} \langle \Phi,\Phi \rangle,
\eea where $\Phi_0=A_0+ \psi_0+ \bar \psi_0.$ 

One can make ${\cal A}\oplus \varPsi\oplus \bar \varPsi$ into a commutative algebra  
by taking as product the bracket $ \langle 
\phantom{n}
,
 \phantom{n}
 \rangle.$ We have
$$\langle {\cal A},\varPsi \rangle \subset \varPsi,\quad \langle {A},{\bar \varPsi} \rangle \subset {\bar \varPsi},\quad \langle {\varPsi},\bar \varPsi \rangle \subset {\cal A},$$ 
all other brackets being zero. 
\section {
Nonlinear equations with quadratic nonlinearity}
Let $\cal V$ be a vector space and let 
\bea \label {qu}
v= v_0+\frac 1 2 \langle v,v \rangle
\eea
be an equation, where $\langle \phantom {n},\phantom {n} \rangle : {\cal V}^2 \to {\cal  V}$ is a bilinear symmetric function, $v_0$ is a given vector, $v$ is an unknown one. 

To solve this equation we introduce a family of functions  $$\langle \ldots . \rangle:
{{\cal V}^m 
  \to {\cal V}},\quad { m\geq 2},$$ defined for $v_1,\ldots ,v_m \in {\cal V}$ by
\bea \label {u}
\langle v_1,\ldots ,v_m \rangle =\frac 1 2 \sum_{r=1}^{m-1} \sum_{1\leq i_1<\ldots < i_r \leq m } \langle \langle v_{i_1},\ldots,v_{i_r} \rangle,
\langle v_1,\ldots,\widehat{v}_{i_1},\ldots,\widehat{v}_{i_r},\ldots,v_{m} \rangle 
 \rangle,
\eea 
where $ \langle v \rangle=v,$ and $\widehat{v}$ means that ${v}$ is omitted.
Since the 
functions $\langle v_1,\ldots ,v_m \rangle$ depend only on the lower order functions $\langle v_1,\ldots ,v_k \rangle$ with $k< m,$ the solution of the system of equation (\ref {u}) can be constructed by induction.For $m=2$ we have an identity, for $m=3$ 
\bea \label {ord}
\langle v_1,v_2,v_{3}\rangle=
\langle \langle v_{1},v_{2}\rangle, v_3 \rangle+\langle \langle v_{1},v_{3}\rangle, v_2 \rangle+\langle  \langle v_2, v_3 \rangle,v_{1} \rangle. 
\eea 
It is easy to prove
that  $ \langle v_1,\ldots ,v_m \rangle$ is an $m-$ linear symmetric function.

For $m\geq 2,1\leq i,j\leq m,$ let  
$$P^m_{ij}: {\cal V}^m\to {\cal  V}^{m-1} $$ be a function 
defined 
by
\bea \label {p}
P^m_{ij}( v_1,\ldots ,v_m ) = 
( \langle v_i,{v}_{j}\rangle ,v_1 ,\ldots,\widehat{v}_{i},\ldots,\widehat{v}_{j},\ldots,v_{m} ). 
\eea 
If $v \in {\cal V}$ is given by  
\bea \label {v}  v = P^2_{12}P^3_{
i_{m-2}j_{m-2}
}\ldots P^{m-1}_
{i_2j_2}
P^m_{i_1j_1}
(v_1,\ldots ,v_m )
\eea 
for some $ (i_1j_1),\ldots,(i_{m-2}j_{m-2}),$ we say that $v$ is a descendant of $(v_1,\ldots ,v_m ).
$

From (\ref{ord}) it follows that $\langle v_1,v_2,v_{3}\rangle$ is given by the sum of all the descendants of its arguments.
The same is true for $\langle v_1,v_2 \rangle.$ 
Assume that $\langle v_1,\ldots,v_k \rangle,$ $ k<m, $ is given by the sum of all the descendants of  $(v_1,\ldots,v_k ).$
Each descendant of  $(v_1,\ldots ,v_m )$ can be written as
\bea \label{dec}\langle r(v_{I}),
s(v_{J})\rangle,\eea 
where $I=(i_1,\ldots ,i_k)$ and $J=(j_1,\ldots ,j_l)$ are increasing multi-indexes\footnote{The multi-index $I=(i_1,\ldots ,i_n)$ is said to be increasing if $i_1< \ldots < i_n.$ },\\ $I\cup J=(1,\ldots, m),$ $r(v_{I})$ and  $s(v_{J})$ are some descendants of  $v_{I}=(v_{i_1},\ldots ,v_{i_k}) $ and $v_{J}=(v_{j_1},\ldots ,v_{j_l}),$ respectively.
It is easy to verify that $I\cap J=\emptyset . $
Summing  all the functions (\ref {dec}), we get the right-hand side of (\ref {u}).
We have thus proved that $\langle v_1,\ldots,v_m \rangle$  is given by the sum of all the descendants of  $(v_1,\ldots ,v_m ).$

Each  descendant can be represented by a diagram. In this diagram an element of  $\cal V$ is represented by the line segment $\feyn {f}$\,\,. A product $\langle v_i,{v}_{j}\rangle$ is represended by the vertex joining the line segments for     $ v_i,{v}_{j}$ and $\langle v_i,{v}_{j}\rangle.$ 
 The graph for $P^m_{ij}(v_1,\ldots,v_m)$ (\ref {p}) is  depicted in 
Figure 1. Here the points labeled  by $1,\ldots, m$ represent the ends of the lines for $v_1,\ldots ,v_m.$  Using this prescription, one can consecutively draw the diagrams for 
$P^m_{i_1j_1}(v_1,\ldots ,v_m ), \\
P^{m-1}_
{i_2j_2}P^m_{i_1j_1}(v_1,\ldots,v_{m}),$ $\ldots,v $ (\ref {v}).The diagram for $v$ has $m-1$ vertices and $m+1$ external lines. 
The auxiliary  points  $1,\ldots, m$ are removed.

\bigskip
\begin{center}
\begin{fmffile}{graph501}
{
\begin{fmfgraph*}(90,70)
\fmfpen {thin}
\fmflabel{$\langle v_i,v_j\rangle $}{o1}
\fmfleft{i1,i2,i3,i4,i5,i6,i7}
\fmfdot{i1,i2,i4,i6,i7}
\fmflabel{$1$}{i1}
\fmflabel{$i$}{i3}
\fmflabel{$j$}{i5}
\fmflabel{$m$}{i7}
\fmf{plain,label=$v_i$,label.side=right}{i3,v1}
\fmf{plain,label=$v_j$,label.side=left}{i5,v1}
\fmf{plain,$\langle v_i,v_j\rangle $,label.side=right}{v1,o1}
\fmfright{o1}
\fmfdot{i1,i2,i3,i4,i5,i6,i7}
\end{fmfgraph*}} 
\end{fmffile}
\end{center}
\bigskip
\begin{center}
Figure 1. Diagram for $P^m_{ij}(v_1,\ldots,v_m).$
\end{center}
\bigskip

For $v_1=\ldots =v_m= v_0$ equation (\ref {u}) takes the form
\bea \label {orr}
\langle v_0^m \rangle =\frac 1 2 \sum_{r=1}^{m-1}\frac {m!} {r!(m-r)!} \langle \langle v_{0}^r \rangle,
\langle v_0^{m-r}\rangle \rangle,\quad \langle v_0^r \rangle= \langle \underbrace{v_0,\ldots ,v_0}_{r 
}\rangle.
\eea 

We claim that 
\bea \label {orro}
v= \langle e^{v_0} \rangle = \sum_{m=0}^\infty \frac {1} {m!}\langle v_0^m \rangle, \quad \langle v_0^0 \rangle=0,
\eea
is a solution of equation (\ref {qu}). Indeed, substituting (\ref {orro}) in (\ref {qu}), we get 
\bean
\sum_{m=2}^\infty \frac {1} {m!}\langle v_0^m \rangle
=\frac {1} {2}\sum_{m=2}^\infty \sum_{r=1}^{m-1} \frac {1} {r!(m-r)!}
\langle \langle v_0^r \rangle,\langle v_0^{m-r} \rangle \rangle
.
\eean
To conclude the proof, it remains to use  (\ref {orr}).

For $ {\cal V} = \mathbb R,$ and $\langle v,v \rangle=v^2$ equation $(\ref {qu})$ takes the form 
$(1/2) v^2 -v+ v_0=0.$
From this it follows that     
$\langle e^{v_0} \rangle = 1-\sqrt{1-2v_0},$
or equivalently, $\langle v_0^m \rangle=b_m v_0^m,$ where  $b_0=0,b_1=1,$ $b_m=1\cdot3\cdot\ldots\cdot(2m-3), m\geq 2.$ 
Therefore the number of the descendants of $(v_1,\ldots,v_m )$ is
$b_m.$
\section {A solution of the electrodynamics equations 
}
In spinor electrodynamics $ {\cal V} = {\cal A}\oplus\varPsi \oplus \bar \varPsi.$
Combining (\ref {qu0}),(\ref {qu}) and (\ref {orro}), we get   
\begin {gather}
A = P_{\cal A} \langle e^{A_0+\psi_0+\bar \psi_0}\rangle,
\notag \\
\label {pot}
\psi = P_{\varPsi} \langle e^{A_0+\psi_0+\bar \psi_0}\rangle, \qquad \bar \psi = P_{\bar \varPsi} \langle e^{A_0+\psi_0+\bar \psi_0}\rangle, 
\end{gather}
where $P_{\cal A},P_{\varPsi},$ and $P_{\bar \varPsi}$ are projectors on the spaces ${\cal A}, \varPsi,$ and $\bar \varPsi $
respectively. 
The fields $A_0,\psi_0$ and $\bar \psi_0$  mutually commute 
under 
the bracket $\langle \ldots . \rangle,$ and therefore
\bea \label {pots}
\langle e^{A_0+\psi_0+\bar \psi_0}\rangle =\langle e^{A_0}e^{\psi_0}e^{\bar \psi_0}\rangle,
\eea
where
\bean
\langle e^{A_0}e^{\psi_0}e^{\bar \psi_0}\rangle=
\sum_{p,q,r=0}^
\infty 
\frac 1 {p!q!r!} 
\langle 
A_0^p,
\psi_0^q, 
\bar \psi_0^r\rangle.
\eean
The function $\langle 
A_0^p,
\psi_0^q, 
\bar \psi_0^r\rangle$ is defined as $0$ if $p=q=r=0,$ and otherwise
\begin{equation*}\label{curru}
\langle 
A_0^p,
\psi_0^q, 
\bar \psi_0^r\rangle=
\langle 
\underbrace{A_0,\ldots,A_0}_{p},\underbrace{\psi_0,\ldots,\psi_0}_{q},\underbrace{\bar \psi_0,\ldots,\bar \psi_0}_{r}\rangle.
\end{equation*}

The system of equations (\ref {b}) possesses an $U(1)$ global symmetry: 
 \begin{equation*}\label{oaoa}
\psi(x) \to \psi'(x)=e^{i\theta}\psi(x),\quad \bar \psi(x) \to \bar \psi'(x)=e^{-i\theta}\bar\psi(x).
\end{equation*}
This implies
\begin{equation*}\label{currm}
P_{\cal A}\langle A_0^p,\psi_0^q, 
\bar \psi_0^r\rangle=
\begin {cases} 
\langle A_0^p,\psi_0^q, \bar \psi_0^r\rangle,&\text {if $q=r,$}\\
0,& \text {otherwise},
 \end{cases}
\end{equation*}

\begin{equation*}\label{currc}
\phantom{a}P_{\varPsi}\langle A_0^p,\psi_0^q, 
\bar \psi_0^r\rangle =
\begin {cases} 
\langle A_0^p,\psi_0^q,\bar \psi_0^r\rangle,&\text {if $q=r+1,$}\\
0,& \text {otherwise},
 \end{cases}
\end{equation*}

\begin{equation*}\label{currp}
\phantom{a}P_{\bar \varPsi}\langle A_0^p,\psi_0^q, 
\bar \psi_0^r\rangle =
\begin {cases} 
\langle A_0^p,\psi_0^q,\bar \psi_0^r\rangle,&\text {if $r=q+1,$}\\
0,& \text {otherwise}.
\end{cases}
\end{equation*}

Using these relations, (\ref {pot}) and  (\ref {pots}), we get
\begin{gather}
{A}=
\sum_{p,q=0}^
\infty 
\frac 1 {p!(q!)^2} 
\langle 
A_0^p,
\psi_0^q, 
\bar \psi_0^q\rangle,
\notag
\\
\psi=
\sum_{p,q=0}^
\infty 
\frac 1 {p!(q+1)!q!} 
\langle 
A_0^p,
\psi_0^{q+1}, 
\bar \psi_0^q\rangle,\notag \\
\Bar \psi=\sum_{p,q=0}^
\infty 
\frac 1 {p!q!((q+1)!} 
\langle 
A_0^p,
\psi_0^{q}, 
\bar \psi_0^{q+1}
\rangle.\notag
\end{gather}

Drawing diagrams in electrodynamics, we depict the  lines  accociated with elements of ${\cal A},\varPsi,$ and $ \bar \varPsi$ by $\feyn {g}$\,\,, $\feyn {fA}$\,\, and \, $\feyn {fV}$\,\,, respectively.
In Figure~2 we show the diagram for $P^m_{ij}(v_1,\ldots, v_m),$ where $v_i=\psi_i \in \varPsi,v_j=\bar \psi_j\in \bar \varPsi.$ 
\bigskip
\begin{center}
\begin{fmffile}{graph502}
\begin{fmfgraph*}(90,70)
\fmfpen {thin}
\fmflabel{$\langle \psi_i,\bar \psi_j\rangle $}{o1}
\fmfleft{i1,i2,i3,i4,i5,i6,i7}
\fmfdot{i1,i2,i4,i6,i7}
\fmflabel{$1$}{i1}
\fmflabel{$i$}{i3}
\fmflabel{$j$}{i5}
\fmflabel{$m$}{i7}
\fmf{fermion, label=$\psi_i$,label.side=right}{i3,v1}
\fmf{fermion, label=$\bar \psi_j$,label.side=right}{v1,i5}
\fmf{photon}{v1,o1}
\fmfright{o1}
\fmfdot{i1,i2,i3,i4,i5,i6,i7}
\end{fmfgraph*}
\end{fmffile}
\end{center}
\bigskip
\begin{center}
Figure 2. Diagram for $P^m_{ij}(v_1,\ldots,\psi_i,\ldots, \bar \psi_j,\ldots, v_m).$
\end{center}
\bigskip

For example, the $O(e^2)$ contribution in $A$ is given by  
\bean
\langle 
A_0,
\psi_0, 
\bar \psi_0\rangle
=
\langle\langle A_0, \psi_0\rangle, \Bar \psi_0\rangle+
\langle\langle A_0, \Bar \psi_0\rangle,  \psi_0\rangle. 
\eean
Recall that the product $ \langle \phantom{n}, \phantom{n} \rangle$ includes a factor of \,$e.$
The diagram for $ \langle\langle A_0, \psi_0\rangle, \Bar \psi_0\rangle $ is depicted in Figure 3.

\bigskip
\begin{center}
\begin{fmffile}{graph503}
{\begin{fmfgraph*}(90,70)
\fmfpen{thin}
\fmflabel{$\langle\langle A_0, \psi_0\rangle, \Bar \psi_0\rangle $}{o1}
\fmfleft{i1,i2,i3
}
\fmflabel{$\psi_0$}{i1}
\fmflabel{$A_0$}{i2}
\fmflabel{$\bar \psi_0$}{i3}
\fmf{photon}{i2,v1}
\fmf{fermion}{i1,v1,v2,i3}
\fmf{photon}{v2,o1}
\fmfright{o1}
\end{fmfgraph*}}
\end{fmffile}
\end{center}
\bigskip
\begin{center}
Figure 3. Diagram for $\langle\langle A_0, \psi_0\rangle, \Bar \psi_0\rangle$
\end{center}

\end{document}